\documentclass[prl,aps,amsmath,twocolumn,amsfonts,showpacs,superscriptaddress]{revtex4}
\usepackage{graphicx}

\def \be#1\ee {\begin{equation}#1\end{equation}}
\def \bea#1\eea {\begin{eqnarray}#1\end{eqnarray}}

\def\str{\mathop{\rm str}}
\newcommand{\corr}[1]{\langle #1\rangle}
\renewcommand{\Re}{\mathop{\rm Re}}
\renewcommand{\Im}{\mathop{\rm Im}}
\newcommand{\eps}{\varepsilon}

\def\br{{\bf r}}

\def\mls{\Delta}
\def\lambdaB{\lambda_1}
\def\lambdaF{\lambda}

\begin{document}

\title{Local correlations of different eigenfunctions in a disordered wire}

\author{M. A. Skvortsov}
\affiliation{L. D. Landau Institute for Theoretical Physics, Moscow 119334, Russia}

\author{P. M. Ostrovsky}
\affiliation{Institut f\"ur Nanotechnologie, Forschungszentrum Karlsruhe,
76021 Karlsruhe, Germany}
\affiliation{L. D. Landau Institute for Theoretical Physics, Moscow 119334, Russia}

\date{July 3, 2006} % base on ldos-paper26.tex

\begin{abstract}
We calculate the correlator of the local density of states
$\corr{\rho_{\eps}(\br_1)\rho_{\eps+\omega}(\br_2)}$
in quasi-one-dimensional disordered wires in a magnetic field,
assuming that $|\br_1-\br_2|$ is much smaller than
the localization length.
This amounts to finding the zero mode of the transfer-matrix
Hamiltonian for the supersymmetric $\sigma$-model,
which is done exactly
by the mapping to the three-dimensional Coulomb problem.
Both the regimes of level repulsion and level attraction are obtained,
depending on $|\br_1-\br_2|$.
We demonstrate that the correlations of different eigenfunctions
in the quasi-one-dimensional and strictly one-dimensional cases
are dissimilar.
\end{abstract}


\maketitle

Modern physics knows a number of paradigmatic models
which describe properties of various, seemingly unrelated, phenomena.
Anderson localization in disordered systems with
one-dimensional geometry is one of such models.
Historically, it played an important role in setting down
the concept of quantum localization.
The model of a one-dimensional (1D) chain
was the first example
of the system where localization takes place
at arbitrary weak disorder \cite{MottTwose61,GertsenshteinVasilev1959}.
Later on Thouless \cite{Thouless77} argued that all the states
are localized also in disordered {\em quasi-one-dimensional}\/ (Q1D) systems,
supporting a large number $N\gg1$ of propagating transverse modes.

During the last three decades it had been realized that
many problems of condensed matter physics and quantum chaos
can be mapped onto Q1D localization.
The most natural example is particle propagation in a disordered wire,
both in the limit of weak \cite{Efetov83} and strong
(granular) \cite{IWZ} disorder.
Besides that, the problem of random banded matrices with a large bandwidth
can be mapped \cite{FM91}
onto the same model. The problem of the quantum $\delta$-kicked rotor
\cite{Casati79,Izrailev90} whose evolution operator looks like
a quasi-random banded matrix also belongs \cite{Fishman,AZ96}
to the Q1D universality class. Recently it was argued
\cite{SBK04} that dynamic localization in quantum dots which can
be described by time-dependent random matrices is equivalent to
Q1D localization as well.

Despite the fact that mathematical description of strictly 1D
disordered systems is rather involved
\cite{Berezinsky73,BerezinskyGorkov79},
our knowledge about them is quite complete.
In this case, the localization length $\xi_{\text{1D}}$ coincides
with the mean free path $l$.
The low-frequency conductivity follows the Mott-Berezinsky law:
$\Re\sigma(\omega)\propto\omega^2\ln^2(1/\omega\tau)$
\cite{Mott68,Berezinsky73}, where $\tau$ is the elastic time.
The correlation function of the local density of states (LDOS)
was considered by Gor'kov, Dorokhov and Prigara \cite{GDP83}
who have shown that in the limit of small energy separation,
$\omega\tau\ll1$, the eigenstates
are uncorrelated at $r\gg \xi_{\text{1D}}\ln(1/\omega\tau)$,
exhibit nearly perfect level repulsion at
$\xi_{\text{1D}}\ll r\ll \xi_{\text{1D}}\ln(1/\omega\tau)$,
with a tendency to level attraction at $r\ll \xi_{\text{1D}}$
(for a discussion of these results using
Mott's arguments, see Ref.~\cite{SI87}).

In the Q1D case, the localization length $\xi\sim Nl$ \cite{Dorokhov}
is much larger than the mean free path, that allows to formulate
the problem of Q1D localization on the language
of the nonlinear supersymmetric $\sigma$-model \cite{Efetov83,Efetov-book}
describing the physics of interacting diffusive modes.
The latter problem can be solved
using the transfer-matrix technique introduced by Efetov and Larkin \cite{EL83}.
The idea of the method is to reduce the evaluation of the functional integral
over the superfield $Q(x)$ to the solution of a differential equation,
in analogy with constructing the Schr\"odinger equation from the
Feynman path integral.
However the resulting transfer-matrix equation at finite frequencies $\omega$
appears to be too complicated (even in the simplest unitary case),
so that no results about correlations of {\em different}\/ eigenfunctions
had ever been obtained in the nonperturbative
localized regime, $\omega<\Delta_\xi$,
where $\Delta_\xi$ is the level spacing at the scale
of the localization length.
Instead, one can solve the transfer-matrix equation
in the limit of $\omega\to +i0$ thus extracting information about statistics
of {\em the same}\/ wave function. In this way one can find
the localization length for various symmetry classes \cite{EL83}
and calculate the whole distribution
function of a certain eigenfunction's intensity $|\psi_n(\br)|^2$
\cite{Mirlin-review}.
Similarly, the low-frequency behavior of the dissipative conductivity
in the Q1D case is not known up to now.

Thus we see that our knowledge about Q1D localization
is far from being complete.
This is a challenge since it is quasi- rather than strictly
one-dimensional localization, which is important for various
applications.
The main technical problem lies
in the complicated form of the transfer-matrix equations
which did not allow to get any results concerning information
about different eigenfunctions
in the localized regime, $\omega<\Delta_\xi$.

In this Letter we make the first step towards the theory
of Q1D localization {\em at finite frequencies}\/
by calculating the correlation function
%($\nu$ is the three-dimensional (3D) density of states
%for spinless particles)
of the LDOS
\be
  R(\omega; \br_1, \br_2)
  =
  \nu^{-2}\corr{\rho_{\eps}(\br_1)\rho_{\eps+\omega}(\br_2)}
\label{R-def}
\ee
in a thick disordered wire in a magnetic field (unitary symmetry);
$\rho_\eps(\br) = \sum_n |\psi_n(\br)|^2 \delta(\eps-\eps_n)$,
and $\nu$ is the three-dimensional (3D) density of states
for spinless particles.
In the present paper
we restrict ourselves to the limit of sufficiently small
spatial separations $|\br_1 - \br_2|\ll\xi$,
when only the zero mode of the
transfer-matrix equation is relevant. This zero mode is found
exactly using the mapping onto
%%the Green function of
an auxiliary 3D Coulomb problem.
The explicit expression (\ref{Psi-result})
for the zero-mode wave function
is the main technical achievement
of this Letter.

\begin{figure}
\begin{center}
\includegraphics{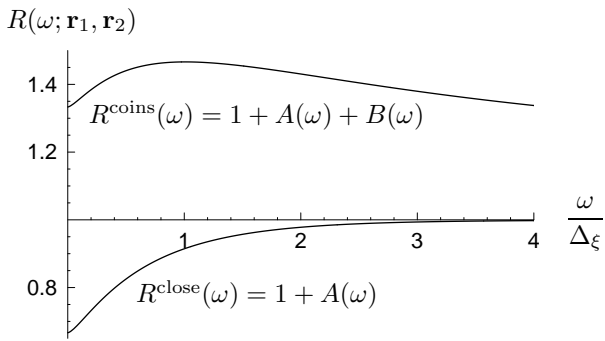}
\end{center}
%\vskip -3mm
\caption{The LDOS correlation function vs.\ $\omega/\mls_\xi$:
at coincident points, $R^{\text{coins}}(\omega)$, and at
$k_F^{-1}\ll|\br_1-\br_2|\ll\xi$, $R^{\text{close}}(\omega)$.
}
\label{F:results}
\end{figure}

Our result for the LDOS correlation function
in the limit $|\br_1 - \br_2|\ll\xi$
has the form
\be
  R(\omega; \br_1, \br_2)
  = 1 + A(\omega) + k(\br_1, \br_2) B(\omega).
\label{R-result}
\ee
The factor $k(\br_1, \br_2) = \corr{\Im G^R(\br_1,\br_2)}^2/(\pi \nu)^2$
accounts for Friedel oscillations \cite{BlanterMirlin97,Mirlin-review}.
Its form is not universal and depends on
the position of the points $\br_1$ and $\br_2$
with respect to the boundaries of the wire.
$k(\br_1, \br_2)$ is equal to 1 at coincident points
and decays fast as $|\br_1-\br_2|$ exceeds the Fermi wavelength $2\pi/k_F$.

The universal functions $A(\omega)$ and $B(\omega)$
determined by slow diffusive modes
are insensitive to a particular geometry of the wire.
They depend on the single dimensionless
parameter $\omega/\Delta_\xi$,
where $\Delta_\xi=(4\pi^2D\nu_1^2)^{-1}$
is the level spacing within the localization region
($D$ is the diffusion coefficient, $\nu_1={\cal A}\, \nu$,
and ${\cal A}\sim Nk_F^{-2}$ is the wire cross-section).
The monotonous functions $A(\omega)$ and $B(\omega)$ are
given explicitly by Eq.~(\ref{A-res}) and (\ref{B-res}).
The function $A(\omega)$ is negative with $A(\infty)=0$ and $A(0)=-1/3$,
whereas $B(\omega)$ is positive with $B(\infty)=0$ and $B(0)=2/3$.

{\em At coincident points}, the correlator
$R^{\text{coins}}(\omega) = 1 + A(\omega) + B(\omega)$
is always larger than 1 signaling short-scale level attraction
(top curve in Fig.~\ref{F:results}).
It starts with
$R^{\text{coins}}(0)=4/3$ in the deeply localized regime,
has a maximum in the crossover region $\omega\sim\Delta_\xi$,
and slowly (as $\sqrt{\Delta_\xi/\omega}$) reaches
the uncorrelated limit $R(\infty)=1$ in the metallic regime ($\omega\gg\Delta_\xi$).
This behavior should be contrasted with the $\omega$-independence of the same correlator
$R^{\text{coins}}_{\text{1D}}(\omega)=1$ for strictly 1D chains
\cite{GDP83}.

{\em For sufficiently close but different points},
$k_F^{-1}\ll|\br_1-\br_2|\ll\xi$,
the correlation function (\ref{R-result}) is nearly
independent of $|\br_1-\br_2|$ and equal to
$R^{\text{close}}(\omega) = 1 + A(\omega)$
which is smaller than 1 indicating level repulsion
(bottom curve in Fig.~\ref{F:results}).
However, this repulsion is weak enough, $R^{\text{close}}(0)=2/3$,
compared to the
perfect level repulsion, $R(0)=0$, in the random matrix theory.
Surprisingly, this value of 2/3 exactly coincides with
the result \cite{GDP83} obtained for the strictly 1D geometry
for $\omega\to0$ in the equivalent limit $k_F^{-1}\ll r\ll \xi_{\text{1D}}$.

%%%%%%%%%%%%%%%%%%%%%%%%%%%%%%%%%%%%%%%%%%%%%%%%%%%%%%%%%%%%%%%%%%%%%%%%%%%%%%
%%%%%%%%%%%%%%%%%%%%%%%%%%%%%%%%%%%%%%%%%%%%%%%%%%%%%%%%%%%%%%%%%%%%%%%%%%%%%%
%%%%%%%%%%%%%%%%%%%%%%%%%%%%%%%%%%%%%%%%%%%%%%%%%%%%%%%%%%%%%%%%%%%%%%%%%%%%%%

We start the technical section of this Letter
by representing the correlation function (\ref{R-def})
in terms of the retarded and advanced Green functions:
\be
   R(\omega; \br_1, \br_2)
   = \frac{1}{2}
   + \frac{\Re \corr{G^R_{\eps+\omega}(\br_1, \br_1) G^A_\eps(\br_2, \br_2)}}
     {2\pi^2\nu^2}
      .
\label{Req1}
\ee
Then we write $G^R$ and $G^A$ as the integrals
over superfields and perform the standard sequence of steps
leading to the Q1D unitary $\sigma$-model \cite{Efetov-book}
formulated in terms of the
supermatrix field $Q$ acting in the direct product of the
Fermi--Bose (FB) and Retarded--Advanced (RA) spaces.
The resulting expression has the form
\be
   R(\omega; \br_1, \br_2)
   = \frac{1}{2}
   - \frac{1}{2} \Re \int P[Q] e^{-S[Q]} DQ(x) ,
\label{R-via-Q-S}
\ee
with the usual diffusive action and the preexponent:
\begin{gather}
  S[Q]
  = \frac{\pi\nu_1}{4} \str \int
  \left[
    D (\nabla Q(x))^2 + 2i \omega \Lambda Q(x)
  \right] dx ,
\label{S[Q]}
\\[5pt]
  P[Q]
  = Q^{\text{RR}}_{\text{BB}}(x_1) Q^{\text{AA}}_{\text{BB}}(x_1)
  + k(\br_1, \br_2) \, Q^{\text{RA}}_{\text{BB}}(x_1) Q^{\text{AR}}_{\text{BB}}(x_1) ,
\label{P(Q)}
\end{gather}
where $\str(\dots)$ is the supertrace,
$\Lambda=\mathop{\rm diag}(1,-1)$ is a matrix
in the RA space, and $Q^{ab}_{\alpha\beta}$ denotes
a single element of the $Q$ matrix.
In deriving $P[Q]$
we have assumed that the field $Q$ does not
fluctuate between the points $x_1$ and $x_2$,
which is true provided that
$|\br_1 - \br_2|\ll\min(\xi,\xi\sqrt{\Delta_\xi/\omega}$).
As we will be mainly concerned with the localized regime,
$\omega<\Delta_\xi$, we will refer to this relation as $|\br_1 - \br_2|\ll\xi$.

Since the preexponent (\ref{P(Q)}) depends only on $Q$
in the single point $x_1$,
the functional integral (\ref{R-via-Q-S}) can be written
as the integral over the single supermatrix ${\cal Q}=Q(x_1)$:
$\int P[Q] e^{-S[Q]} DQ(x)
= \int P({\cal Q}) \Psi^2({\cal Q})\, d{\cal Q}$,
where $\Psi({\cal Q}) = \int_{Q(x_1)={\cal Q}} e^{-S[Q]} DQ(x\geq x_1)$.
In the transfer-matrix approach, the function $\Psi({\cal Q})$
is obtained as the zero-energy solution of an appropriate
supermatrix Hamiltonian \cite{EL83}.
In Efetov's parameterization \cite{Efetov83,Efetov-book},
$\Psi({\cal Q})=\Psi(\lambdaF,\lambdaB)$ depends only on two
variables: $\lambdaF$ and $\lambdaB$, parameterizing the FF
and BB sectors of the supermatrix ${\cal Q}$, respectively.
Taking the integrals over the other variables of the parameterization
we obtain the following expressions for the functions
defined in Eq.~(\ref{R-result}):
\begin{gather}
   A(\omega)
   =
   \frac12
   \Re \int
   \Psi^2(\lambdaF,\lambdaB) \,
   d\lambdaF \, d\lambdaB ,
\label{A-def}
\\
   B(\omega)
   =
   \frac12
   \Re \int
   \frac{\lambdaB+\lambdaF}{\lambdaB-\lambdaF} \,
   \Psi^2(\lambdaF,\lambdaB) \,
   d\lambdaF \, d\lambdaB ,
\label{B-def}
\end{gather}
where the integrals are taken over the strip $-1\leq\lambdaF\leq1$
and $\lambdaB\geq1$.

The function $\Psi(\lambdaF,\lambdaB)$ is the zero mode ($H\Psi=0$)
of the transfer-matrix Hamiltonian \cite{Efetov-book,EL83}:
\begin{multline}
  H = -\frac{(\lambdaB-\lambdaF)^2}2
  \biggl[
    \frac{\partial}{\partial\lambdaF}
    \frac{1-\lambdaF^2}{(\lambdaB-\lambdaF)^2}
    \frac{\partial}{\partial\lambdaF}
\\ {}
  +
    \frac{\partial}{\partial\lambdaB}
    \frac{\lambdaB^2-1}{(\lambdaB-\lambdaF)^2}
    \frac{\partial}{\partial\lambdaB}
  \biggr]
  - \frac{i\omega}{4\Delta_\xi} (\lambdaB-\lambdaF)
\label{H}
\end{multline}
with the boundary condition $\Psi(1,1)=1$.
% due to supersymmetry.

The Hamiltonian (\ref{H}) without the potential term
$\propto(\lambdaB-\lambdaF)$ had been studied in detail \cite{MMZ94,Rejaei96}.
The zero mode
in the deeply localized
regime, $\omega\ll\Delta_\xi$, had been obtained
%by Efetov and Larkin~\cite{EL83}.
in Ref.~\cite{EL83}.
In this limit it depends only on the noncompact bosonic variable $\lambdaB$
and has the form
$\Psi(\lambdaF,\lambdaB) \approx
\sqrt{-2i(\omega/\Delta_\xi)\lambdaB}\,
K_1(\sqrt{-2i(\omega/\Delta_\xi)\lambdaB})$,
with $K_1$ being the MacDonald function.

\begin{figure}
\begin{center}
\includegraphics{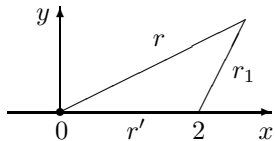}
\end{center}
\vskip -4mm
\caption{To the construction of elliptic coordinates.}
\label{F:elliptic}
\end{figure}

Now we show that the zero mode of the Hamiltonian~(\ref{H})
can be found exactly for an arbitrary $\omega$.
The region of variation of the parameters $\lambdaF$ and $\lambdaB$
suggests us to consider them as elliptic coordinates on the half-plane
$y>0$ of an auxiliary plane $(x,y)$:
$\lambdaF = (r - r_1)/2$ and $\lambdaB = (r + r_1)/2$,
where $r = \sqrt{x^2 + y^2}$ and $r_1 = \sqrt{(x-2)^2 + y^2}$,
see Fig.~\ref{F:elliptic}.
Changing variables from $(\lambdaF,\lambdaB)$ to $(x,y)$
and introducing a new function $\Phi$:
$\Psi = r_1\Phi$,
we obtain
\be
  H\Psi
  = -\frac{r_1^2r}2
  \left[
    \frac{\partial^2}{\partial x^2}
  + \frac{\partial^2}{\partial y^2}
  + \frac 1y \frac{\partial}{\partial y}
  + \frac{i\omega}{2\Delta_\xi r}
  \right] \Phi
  = 0.
\ee
The differential part of this operator resembles the Laplace operator
written in cylindrical coordinates.
Rotating the upper half-plane around the $x$ axis by
the angle $\varphi$ we span the whole 3D space (the function
$\Phi$ is evidently independent of $\varphi$).
The initial condition $\Psi(1,1)=1$
transforms to $\Phi(r_1 \to 0) = 1/r_1$,
which can be easily incorporated into the equation for $\Phi$:
\be
  \left[
    \nabla^2_\br + i\omega/2\Delta_\xi r
  \right] \Phi(\br)
  = - 4\pi\delta^{(3)}(\br-\br') ,
\label{H-Coulomb}
\ee
where $\br'$ is the 3D vector pointing from the origin to the point (2,0,0),
so that $|\br-\br'|=r_1$.
Comparing with the equation
$\{ \nabla_\br^2 - 2\alpha/r + k^2 \} G_k(\br,\br')
   = \delta^{(3)}(\br-\br')$,
which defines the Green function
in the Coulomb potential $\alpha/r$ at energy $k^2/2$,
we immediately conclude that $\Phi=-4\pi G_0(\br,\br')$
is nothing but the zero-energy Green function
in the field of an {\em imaginary}\/ charge
$\alpha=-i\omega/4\Delta_\xi$. To avoid possible confusion
we emphasize that this `zero-energy' refers to the auxiliary
Coulomb problem, while the frequency $\omega$ can be arbitrary.
We also note that it is only the zero mode of the Hamiltonian (\ref{H})
that can be reduced to the Coulomb problem:
for a finite $E$, the equation $H\Psi=E\Psi$ would correspond to the motion
in a non-central field $-i\omega/4\Delta_\xi r-E/rr_1$.

%%%{\em Coulomb Green function.}---%

Surprisingly, the Coulomb Green function has a very simple form
in the real-space representation.
According to the seminal result by Hostler and Pratt \cite{HostlerPratt63},
\begin{multline}
   G_k(\br,\br')
   = \frac{i\Gamma(1+\frac{i\alpha}{k})}{4\pi k|\br-\br'|}
     (\partial_u-\partial_v)
\\{}
  \times
  W_{-i\alpha/k,1/2}(-iku) \, M_{-i\alpha/k,1/2}(-ikv) ,
\label{G-HP}
\end{multline}
where $u = r + r' + |\br-\br'|$, $v = r + r' - |\br-\br'|$,
and $W$ and $M$ are Whittaker's confluent hypergeometric functions.
Taking the $k\to0$ limit of Eq.~(\ref{G-HP})
with the help of the asymptotic formulae
for Whittaker's functions \cite{Bateman},
we obtain the Green function at zero energy:
\be
  G_0(\br,\br')
  = \frac{(\partial_u-\partial_v)
    \sqrt{u} \: K_{1}(2\sqrt{\alpha u})
    \:
    \sqrt{v} \: I_{1}(2\sqrt{\alpha v})}
    {2\pi |\br-\br'|} .
\label{G-HP-0}
\ee

In order to get the function $\Psi(\lambdaF,\lambdaB)$
one has to multiply $\Phi=-4\pi G_0(\br,\br')$ by $r_1$
that cancels the factor $|\br-\br'|$ in the denominator
of Eq.~(\ref{G-HP-0}), and express $u$ and $v$
in terms of $\lambdaF$ and $\lambdaB$:
$u = 2(\lambdaB+1)$, $v = 2(\lambdaF+1)$.
Thus we obtain the explicit form of the zero mode of the
Hamiltonian (\ref{H}) valid for an arbitrary frequency $\omega$:
\be
  \Psi(\lambdaF,\lambdaB)
  = K_0(p) \, q I_1(q) + p K_1(p) \, I_0(q) ,
\label{Psi-result}
\ee
where we have denoted $p = \sqrt{-2i(\omega/\Delta_\xi)(\lambdaB+1)}$
and $q = \sqrt{-2i(\omega/\Delta_\xi)(\lambdaF+1)}$.
The properties of the Bessel functions
ensure proper normalization: $\Psi(1,1)=1$.

The form of Eq.~(\ref{Psi-result}) indicates that
the initial coordinates $\lambdaF$ and $\lambdaB$ were somehow
`more natural' that the auxiliary 3D Coulomb coordinates.
Nevertheless we think than the mapping
to the Coulomb problem is important as it reveals the
high symmetry of the transfer-matrix Hamiltonian (\ref{H})
related to the $O(4)$ symmetry of the Coulomb problem \cite{Fock35,Bander}.

The knowledge of the zero mode $\Psi(\lambda,\lambda_1)$
allows us to calculate $A(\omega)$ and $B(\omega)$
in Eqs.~(\ref{A-def}) and (\ref{B-def}) analytically.
The integration for $A(\omega)$ is done in terms
of indefinite integrals of the product of two Bessel functions.
The integral for $B(\omega)$ is much simplified
in the representation of the 3D Coulomb coordinates:
$B(\omega)=4\pi \Re \beta(\omega)$ with
\be
  \beta(\omega)
  = \int d^3\br \, G_0^2(\br, \br')
  = - \lim_{\br''\to\br'} \lim_{k\to0}
      \frac{\partial G_k(\br'', \br')}{\partial k^2} ,
\label{B}
\ee
which is calculated using the exact expression (\ref{G-HP}).
The results for $A(\omega)$ and $B(\omega)$ read
($\kappa=\sqrt{-4i\omega/\Delta_\xi}$):
\begin{multline}
  A(\omega)
  = (4/3) \Re \Bigl\{
    \kappa^2 \bigl[ I_1^2(\kappa) - I_0(\kappa) I_2(\kappa) \bigr]
\\
{}  \times
             \bigl[ K_1^2(\kappa) - K_0(\kappa) K_2(\kappa) \bigr]
    - I_1^2(\kappa) K_1^2(\kappa)
    \Bigr\} ,
\label{A-res}
\end{multline}
%and
\vskip -7mm
\be
  B(\omega)
  = (4/3) \Re
  \bigl[ I_1(\kappa) K_1(\kappa) + 2I_2(\kappa) K_0(\kappa) \bigr] ,
\label{B-res}
\ee
where we have used the properties of the Bessel functions and
the fact that $\kappa^2$ is imaginary ($\omega$ is real)
in order to simplify the result.
Exact Eqs.~(\ref{A-res}) and (\ref{B-res}) interpolate between
the metallic regime ($\omega\gg\Delta_\xi$), which can be treated
perturbatively by expanding in diffusive modes,
and the nonperturbative localized regime ($\omega\ll\Delta_\xi$).
The asymptotic expressions have the form ($z=\omega/\Delta_\xi$):
\begin{gather}
\!
  A(\omega) =
  \begin{cases}
    \displaystyle
    -1/3 + z^2\ln^2(1/z)/3 + \dots , & z\ll1 ,
    \\
    \displaystyle
    -3/(64z^2) + \dots ,
      & z\gg1 ;
  \end{cases}
\\
  B(\omega) =
  \begin{cases}
    \displaystyle
    2/3 - 2z^2\ln(1/z)/9 + \dots , & z\ll1 ,
    \\
    \displaystyle
    \sqrt{1/2z} + \dots , & z\gg1 .
  \end{cases}
\end{gather}

Now let us compare eigenfunctions' correlations
in the 1D and Q1D geometries.
It is known that the {\em single}\/ eigenfunction statistics
in the two cases are closely related \cite{Mirlin-review}.
Namely, the statistics of the wave function envelopes
are precisely the same while the short-scale oscillations are different.
Having derived the LDOS correlator in the Q1D case, we are now able to
compare correlations of {\em different}\/ eigenfunctions
in the Q1D and 1D cases.
The most striking mismatch between the two problems is seen
in the LDOS correlation function at coincident points: our result
for $R^{\text{coins}}(\omega)$ is always $>1$ with
a nontrivial $\omega$-dependence (top curve in Fig.~\ref{F:results}),
whereas $R^{\text{coins}}_{\text{1D}}(\omega)\equiv1$ for the strictly 1D
geometry \cite{GDP83}.
However since the scales shorter than the Fermi wavelength can
hardly be resolved (and with the results being model-dependent),
it is more instructive to compare $R^{\text{close}}(\omega)$
at spatial separations $|\br_1-\br_2|$ larger than $k_F^{-1}$
and smaller than the localization length ($\xi$ for Q1D, and $l$ for 1D).
For the 1D geometry,
$R^{\text{close}}_{\text{1D}}(\omega)
= 1 - (1/3) \int_0^\infty dt \, t^3 e^{-t}/[(\omega\tau)^2+t^2]$
\cite{BerezinskyGorkov79,GDP83},
which is definitely different from our result,
$R^{\text{close}}(\omega)=1+A(\omega)$.
Though $R^{\text{close}}(0) = R^{\text{close}}_{\text{1D}}(0) = 2/3$,
already the next term in the small-$\omega$ expansion
[$\omega^2\ln^2(\Delta_\xi/\omega)$ vs.\ $\omega^2\ln(1/\omega\tau)$]
demonstrates the difference between
the two problems even in the deeply localized regime.
Thus we conclude that the analogy between 1D and Q1D localization seen
in the correlations of single wave functions, {\em does not extend}\/ to
the correlations of different wave functions.

Finally, we mention that the same approach can be employed
in calculating the density-density correlation function
which determines the kinetic response of the wire.
In the limit $|\br_1-\br_2|\ll\xi$
we get a counterpart of Eq.~(\ref{R-result})
with $1+A(\omega)$ and $B(\omega)$ interchanged:
\begin{multline}
  \corr{\Im G^R_\eps(\br_1, \br_2)
        \Im G^R_{\eps+\omega}(\br_2, \br_1)}
\\ {}
  = (\pi\nu)^2 [k(\br_1, \br_2)(1+A(\omega))+B(\omega)].
\label{GG}
\end{multline}
Note however that Eq.~(\ref{GG}) is not sufficient to
get the conductivity of the wire, which requires
a much more complicated analysis of the limit $|\br_1-\br_2|\sim\xi$
\cite{EL83,Efetov-book}.

To conclude, we obtained the explicit expression
for the zero mode of the supersymmetric transfer-matrix
Hamiltonian of a thick disordered wire in a magnetic field.
This allowed us to get the nonperturbative result for
the pair correlation function of LDOS,
which describes correlations
between different eigenfunctions, in the limit $|\br_1-\br_2|\ll\xi$.
We demonstrate that these correlations are different in the 1D and
Q1D cases.
Generalization to arbitrary $|\br_1-\br_2|$ and other symmetry
classes will be the subject of further studies.

We thank D. M. Basko, M. V. Feigel'man,  I. V. Gornyi, D. A. Ivanov,
I. V. Kolokolov, V. E. Kravtsov and A. D. Mirlin
for fruitful discussions. The research of M.~A.~S. was
supported by the Program ``Quantum Macrophysics''
of the Russian Academy of Sciences,
RFBR under grant No.\ 04-02-16998,
and the Russian Science Support Foundation.


\begin{thebibliography}{99}

\bibitem{MottTwose61}
N. F. Mott and W. D. Twose, Adv. Phys. {\bf 10}, 107 (1961).

\bibitem{GertsenshteinVasilev1959}
M. E. Gertsenshtein and V. B. Vasil'ev,
Teor. Veroyatn. Primen. {\bf 4}, 424 (1959); {\em ibid}. {\bf 5}, 3(E) (1959)
[Theor. Probab. Appl. {\bf 4}, 391 (1959); {\em ibid}. {\bf 5}, 340(E) (1959)].

\bibitem{Thouless77}
D. J. Thouless,
Phys. Rev. Lett. {\bf 39}, 1167 (1977).

\bibitem{Efetov83}
K. B. Efetov, Adv. Phys. {\bf 32}, 53 (1983).

\bibitem{IWZ}
S. Iida, H. A. Weidenm\"uller, J. A. Zuk, Ann. Phys. (NY) 200, 219 (1990).

\bibitem{FM91}
Y. V. Fyodorov and A. D. Mirlin,
Phys. Rev. Lett. {\bf 67}, 2405 (1991).

\bibitem{Casati79}
G. Casati, B. V. Chirikov, J. Ford, and F. M. Izrailev,
Lect. Notes Phys. {\bf 93}, 334 (1979).

\bibitem{Izrailev90}
F. M. Izrailev, Phys. Rep. {\bf 196}, 299 (1990).

\bibitem{Fishman}
S. Fishman {\em et al}., Phys. Rev. Lett. {\bf 49}, 509 (1982);
D. R. Grempel, R. E. Prange, and S. Fishman,
Phys. Rev. A {\bf 29}, 1639 (1984).

\bibitem{AZ96}
A. Altland and M. R. Zirnbauer, Phys. Rev. Lett. {\bf 77}, 4536 (1996).

\bibitem{SBK04}
M. A. Skvortsov, D. M. Basko, V. E. Kravtsov,
Pis'ma v ZhETF {\bf 80}, 60 (2004) [JETP Lett. {\bf 80}, 54 (2004)].

\bibitem{Berezinsky73}
V. L. Berezinsky,
Zh. Eksp. Teor. Fiz. {\bf 65}, 1251 (1973) [Sov. Phys. JETP {\bf 38}, 620 (1974)].

\bibitem{BerezinskyGorkov79}
V. L. Berezinsky and L. P. Gor'kov,
Zh. Eksp. Teor. Fiz. {\bf 77}, 2498 (1979) [Sov. Phys. JETP {\bf 50}, 1209 (1979)].

\bibitem{Mott68}
N. F. Mott, Phylos. Mag. {\bf 17}, 1259 (1968).

\bibitem{GDP83}
L. P. Gor'kov, O. N. Dorokhov, and F. V. Prigara,
Zh. Eksp. Teor. Fiz. {\bf 84}, 1440 (1983) [Sov. Phys. JETP {\bf 57}, 838 (1983)].

\bibitem{SI87}
U. Sivan and Y. Imry, Phys. Rev. B {\bf 35}, 6074 (1987).

\bibitem{Dorokhov}
O. N. Dorokhov, Pis'ma v Zh. Eksp. Teor. Fiz. {\bf 36}, 259 (1982)
[Sov. Phys. JETP Lett. {\bf 36}, 318 (1982)].

\bibitem{Efetov-book}
K. B. Efetov, {\em Supersymmetry in Disorder and Chaos}
(Cambridge University Press, New York, 1997).

\bibitem{EL83}
K. B. Efetov and A. I. Larkin,
Zh. Eksp. Teor. Fiz. {\bf 85}, 764 (1983)
[Sov. Phys. JETP {\bf 58}, 444 (1983)].

\bibitem{Mirlin-review}
A. D. Mirlin, Phys. Rep. {\bf 326}, 259 (2000).

\bibitem{BlanterMirlin97}
Ya. M. Blanter and A. D. Mirlin,
Phys. Rev. E {\bf 55}, 6514 (1997).

\bibitem{Rejaei96}
B. Rejaei, Phys. Rev. B {\bf 53}, R13235 (1996).

\bibitem{MMZ94}
A. D. Mirlin, A. M\"uller-Groeling, M. R. Zirnbauer,
Ann. Phys. (NY) {\bf 236}, 325 (1994).

\bibitem{HostlerPratt63}
L. Hostler and R. H. Pratt, Phys. Rev. Lett. {\bf 10}, 469 (1963).

\bibitem{Bateman}
H. Bateman, A. Erd\'elyi, {\em Higher transcendental functions}, Vol.\ 1
(McGraw-Hill, New York, 1953).

%\bibitem{LL1}
%L. D. Landau and E. M. Lifshits, {\em Mechanics}
%(Pergamon, Oxford, 1978).

\bibitem{Fock35}
V. Fock, Z. Physik {\bf 98}, 145 (1935).

\bibitem{Bander}
M. Bander and C. Itzykson,
Rev. Mod. Phys. {\bf 38}, 320 (1966); {\em ibid}. {\bf 38}, 346 (1966).

%\bibitem{Schwinger64}
%J. Schwinger, J. Math. Phys. {\bf 5}, 1606 (1964).


\end{thebibliography}
\end{document}